\documentclass[11pt]{article}
\usepackage{moriond,graphicx,amsmath}

\def\be{\begin{equation}}
\def\ee{\end{equation}}
\def\bea{\begin{eqnarray}}
\def\eea{\end{eqnarray}}

\begin{document}
\title{Detection of quantum noise from mesoscopic devices with an SIS detector}

\author{\underline{R. Deblock}$^{1,2}$, E. Onac$^1$, L. Gurevich$^1$, L.P. Kouwenhoven$^1$}
\address{$^1$Department of Nanoscience and ERATO Mesoscopic Correlation Project, Delft University of Technology, POBox 5046, 2600 GA Delft, Netherlands\\
$^2$Present address: Laboratoire de Physique des Solides, Université Paris-Sud, Bât. 510, 91
405 Orsay Cedex, France}

\maketitle
\abstracts{
Quantum mechanics can strongly influence the noise properties of mesoscopic devices. To probe this effect we have measured the current fluctuations at high-frequency (5-90 GHz) using a superconductor-insulator-superconductor tunnel junction as an on-chip spectrum analyzer. By coupling this frequency-resolved noise detector to a quantum device we can measure the high-frequency, non-symmetrized noise as demonstrated for a Josephson junction. The same scheme has been used to detect the current fluctuations arising from coherent charge oscillations in a two-level system, a superconducting charge qubit. A narrow band peak was observed in the spectral noise density at the frequency of the coherent charge oscillations \cite{deblock}.}

Measurement of shot-noise has proved to be a powerful probe of electronic properties in mesoscopic devices \cite{buttiker}. It allows to have information on the system that are not accessible by conductance measurement such as correlation between electrons or charge of the current carrier. A very interesting limit is when frequency is of the order or higher than $k_B T$ and $e V$, with $T$ the temperature and $V$ the bias voltage. In this limit one can be sensitive to the zero-point fluctuation of the electromagnetic field or to the internal energy scale of the mesoscopic device under study. Only very few experiments in this regime are available \cite{schoelkopf,koch} due to the difficulty to work in the correct range of frequencies, which is typically between 1 and 100 GHz for mesoscopic devices. We present a scheme to measure the current fluctuations in the frequency range 5 to 80 GHz by using a superconductor-insulator-superconductor (SIS) junction as an on-chip spectrum analyzer.
This technique is used to detect the HF emission of a Josephson junction. In ref. \cite{deblock} this scheme has been successfully applied to measure the quantum noise of a two-level system, a Cooper pair box.

\section{Method}

The idea of the detection is to couple on chip a detector with the system under study and is thus in line with Ref. \cite{aguado,schoelkopf2}. The detector used in this work is a SIS junction, a well established device in astrophysics (where it is often used as a high-frequency (HF) mixer) \cite{tucker}. 

\begin{figure}
	\begin{center}
		\includegraphics[width=16cm]{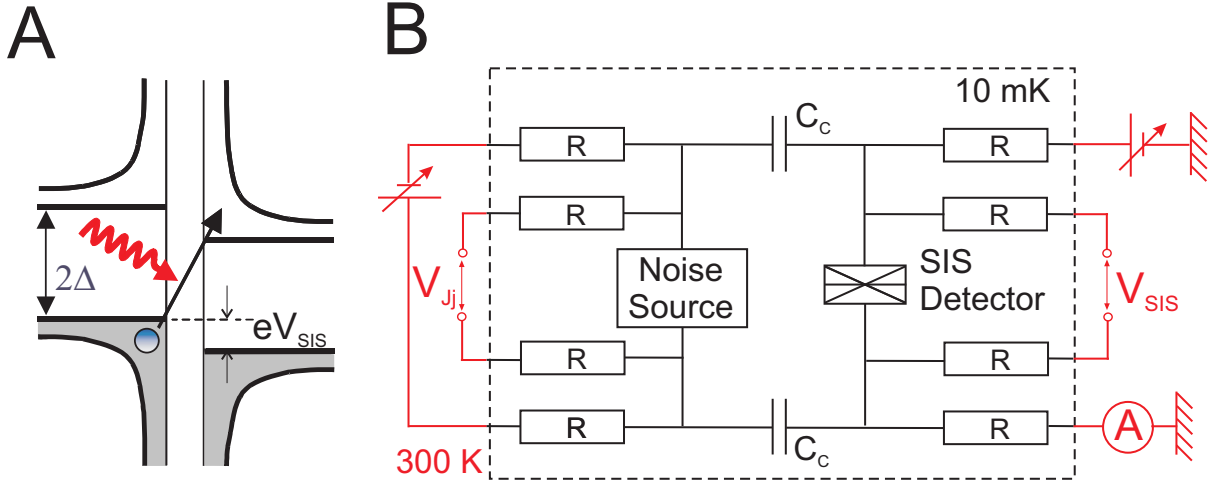}
	\end{center}
	\caption{\textbf{A} Schematic picture of the photon-assisted tunneling through a SIS junction. \textbf{B} Schematic of the coupling circuit between the device (the noise generator) and the SIS detector. The circuit is made with on-chip resistances $R$ and coupling capacitances $C_C$.}
	\label{fig:circuit}
\end{figure}

In the following we consider only the quasi-particle current through this SIS junction. For bias $V_{SIS}$ below $2 \Delta/e$, due to the gap $2 \Delta$ in the density of state of the superconductors, there is no current. However when one has $e |V_{SIS}| > 2 \Delta$ a quasi-particle current can flow and one recovers the normal state resistance of the junction. For $e |V_{SIS}| < 2 \Delta$, if the junction is submitted to photons, a quasi-particle current can exist provided that the photons have enough energy to allow the tunneling of quasi-particles (Fig. \ref{fig:circuit} A). Thus for photons of energy $\hbar \omega$ one has a non-zero photon-assisted tunneling (PAT) current for bias such that $2\Delta -e V_{SIS} < \hbar \omega$. The PAT current carries information on the number and the frequency of the photons reaching the junction. Note that the current for bias higher than the gap is also affected by the photons.

To be more quantitative and take into account the fact the electromagnetic field can have a broad range of frequencies, we consider a SIS junction coupled to an environment characterized by a voltage spectral density $S_{V}(\omega)$. Note that here we consider both negative and positive frequencies. The negative frequencies correspond to an energy flow from the environment to the SIS junction, i.e. the emission part of the spectrum for the environment, whereas the positive frequencies corresponds to energy being absorbed by the environment. This distinction is important at high frequencies \cite{gavish,lesovik}, depending on the detection scheme. 
An SIS junction coupled to an environment has been considered by Ingold and Nazarov \cite{nazarov}. At zero temperature and for $V>0$ the quasiparticle current in an SIS junction reads :
\begin{equation}
	I_{QP}(V)=\int_{0}^{+\infty} d\epsilon \, P(eV-\epsilon) I_{QP,0}\left( \frac{\epsilon}{e} \right)
\label{Iqp}
\end{equation}
with $I_{QP,0} (V)$ the quasi-particle current without noise, $P(\epsilon)=1/2 \pi \hbar \int_{-\infty}^{+\infty} \exp \left[ J(t)+i \epsilon t /\hbar \right] dt$ the probability to exchange energy with the environment, with $J(t)=<\delta\widetilde{\varphi}(t)\delta\widetilde{\varphi}(0)-\left( \delta\widetilde{\varphi}(0) \right)^2>$ the phase-phase correlator with $\delta\widetilde{\varphi}(t)=\frac{e}{\hbar} \int^{t} dt^{'} \delta V(t^{'})$. Note that this treatment relies on the fact that the noise is gaussian. In particular if the device under study has an important third cumulent this treatment may be wrong. For an environment characterized by \textit{non-symmetrized} voltage fluctuations $S_{V}(\omega)$  we get, with $R_K = h/e^2$ the quantum resistance \cite{aguado}:
\begin{equation}
	J(t)=\frac{2 \pi}{\hbar R_K} \int_{-\infty}^{+\infty} d\omega \, S_{V}(\omega) \left[\exp{(-i \omega t)}-1 \right]	
\end{equation}
Assuming $J(t)$ always much smaller than 1, so that $\exp{(J(t))} \approx 1+J(t)$, we deduce :
\begin{equation}
	P(\epsilon)=\left[1-\frac{2 \pi}{\hbar R_K} \int_{-\infty}^{+\infty} d\omega \, S_{V}(\omega) \right] \delta(\epsilon)+\frac{2 \pi}{R_K} \frac{S_{V}(\epsilon/\hbar)}{\epsilon^2} 
\end{equation}
Inserting this expression in eq. \ref{Iqp} we get the photon assisted tunneling current.
\begin{eqnarray}
	I_{PAT}(V)&=&I_{QP}(V)-I_{QP,0}(V) \nonumber \\
	&=& \int_{0}^{+\infty} d\omega \, \left( \frac{e}{\hbar\omega} \right)^2 S_{V}(-\omega) I_{QP,0} \left( V+\frac{\hbar \omega}{e} \right) \nonumber \\
	&+& \int_{0}^{eV} d\omega \, \left( \frac{e}{\hbar\omega} \right)^2 S_{V}(\omega) I_{QP,0} \left( V-\frac{\hbar \omega}{e} \right) - \int_{-\infty}^{+\infty} d\omega \, \left( \frac{e}{\hbar\omega} \right)^2 S_{V}(\omega) I_{QP,0} \left( V \right)
	\label{Ipat}
\end{eqnarray}
The first term corresponds to emission ($\omega < 0$), the second to absorption ($\omega > 0$) and the third one renormalizes the elastic current. An experimentally important case  is when $V < 2 \Delta$. Then $I_{QP,0} \left( V-\frac{\hbar \omega}{e} \right)=0$ and $I_{QP,0} \left( V \right)=0$ so that the latter equation simplifies to :
\begin{equation}
	I_{PAT}(V)= \int_{0}^{+\infty} d\omega \, \left( \frac{e}{\hbar\omega} \right)^2 S_{V}(-\omega) I_{QP,0} \left( V+\frac{\hbar \omega}{e} \right)
	\label{Ipatsimple}
\end{equation}
The detector is then sensitive only to the emission part of the voltage noise spectrum ($\omega<0$). The SIS detectors are sensitive up to frequencies $~2 \Delta/h$. For aluminium junctions, which are used in this work, this is around 100 GHz. Niobium junctions go up to THz.

In order to achieve a good sensitivity one has to provide a good coupling between the device under study and the detector, i.e. the SIS junction, in the frequency range of interest (5-90 GHz for aluminum SIS junctions). We have achieved this by using an on-chip circuitry based on resistances and capacitances (Fig \ref{fig:circuit} B and C). The capacitances (estimated value : 550 fF, area : 80x10 $\mu$m$^2$, 50 nm of SiO) are designed to provide a strong coupling of the detector with the device at high frequency and to allow independent DC biasing of the two part of the circuit. The resistance (value : 2 k$\Omega$, 20 $\mu$m long platinum wires of width 100 nm and thickness 25 nm) are used to isolate the circuit composed of the detector, the capacitances and the device from the external circuit and in particular from the capacitances to the ground provided by the wiring from room temperature to the sample.

From the value of the elements of the on-chip circuitry it is possible to calculate the expected transimpedance $Z(\omega)$ of the coupling circuit, defined as the ratio between the voltage fluctuations across the junction and the current fluctuations through the device at frequency $\omega$.  However this type of calculation is only indicative concerning the transimpedance that we will eventually have in the real experiment. Indeed it is rather complicated to have a correct modelisation of all the component on chip at high frequency. Thus to be able to make quantitative statement concerning the experiment one has to have a way to calibrate the on-chip circuit, i.e. measure $Z(\omega)$. 

\section{High-frequency emission of a Josephson Junction}
In order to calibrate the noise detection with the SIS detector using our coupling circuit we first measure the already well characterized noise generated by a Josephson junction. This device allows us to do detection of a narrow band noise and a broad band noise depending of the bias conditions.

\begin{figure}
	\begin{center}
		\includegraphics[width=7.9cm]{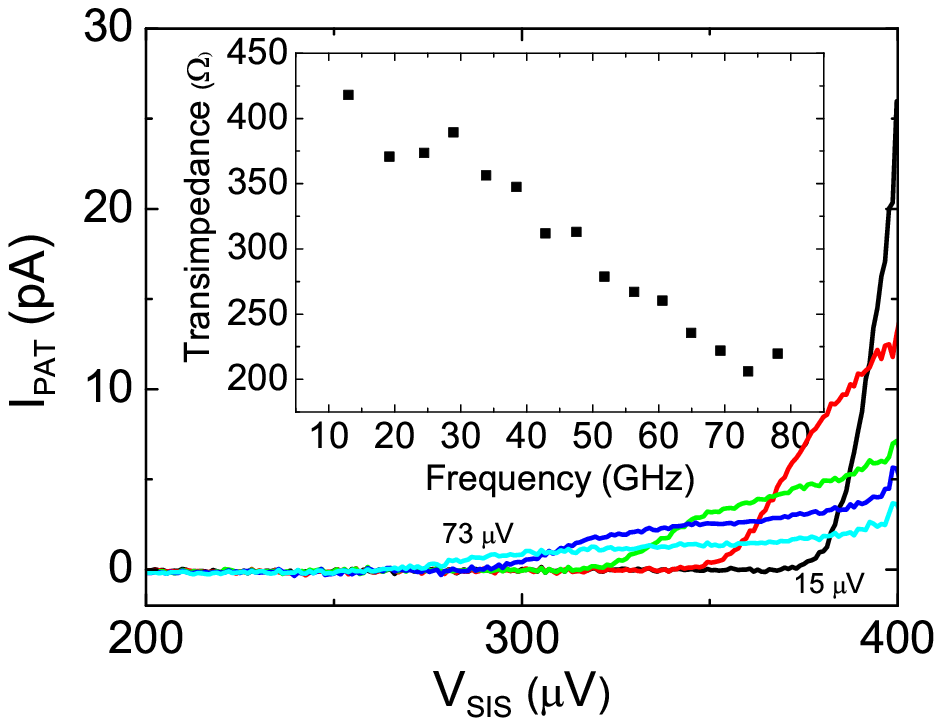}
		\includegraphics[width=7.9cm]{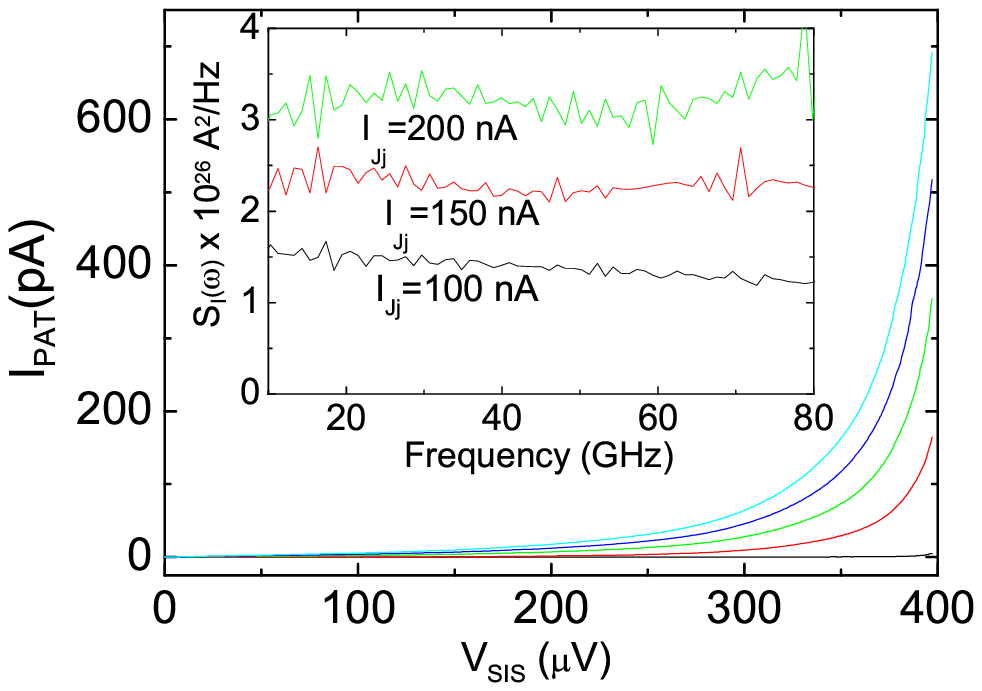}
	\end{center}
	\caption{Left figure : measured PAT current for $V_{Jj}=15, 30, 45, 60, 73 \mu$V for bias $V_{SIS}$ of the detector below $2 \Delta/e = 420 \mu$V. Inset : measured frequency dependence of the transimpedance $|Z(\omega)|$. Right figure : $I_{PAT}$ for quasi-particle current through the Josephson junction $I_{Jj}=10, 50, 100, 150, 200$ nA (from bottom to top). Inset : Current spectral density extracted from the $I_{PAT}$ current for different$I_{Jj}$. It is consistent with the value of the non-symmetyrized noise $e I_{Jj}$.}
	\label{fig:SNJJ}
\end{figure}

When the bias $V_{Jj}$ of the junction is below $2 \Delta$ there is no quasi-particle current but there is a Cooper-pair current characterized by the two relations \cite{tinkham} $I = I_C \sin(\Phi)$ and $d \Phi/dt = 2 e V_{Jj}/\hbar$ with $\Phi$ the superconducting phase difference across the junction. As a consequence for a finite DC bias, the junction is acting as a HF current generator with a AC current of amplitude $I_C= \pi \Delta /(2 e R_N)$ and a frequency $f = 2 e V_{Jj}/ h$ ($R_N$ is the normal state resistance of the junction). The PAT current measured (Fig \ref{fig:SNJJ} left) is then typical of a narrow band noise spectrum. From this and eq. \ref{Ipatsimple} we can deduce $Z(\omega)$ at frequency $2 e V_{Jj}/ h$ (Inset of Fig \ref{fig:SNJJ}). Note that it should be possible in principle to extract the frequency width of the AC Josephson effect. However this is true only if this width is bigger than the width of the transition at $2 \Delta$ in the IV characteristic of the SIS detector. In our case we are in the opposite limit which makes the extraction of the width of the AC Josephson effect impossible.

When $V_{Jj} > 2 \Delta$ there is a quasi-particle current $I$ with an associated shot-noise $S_I(\omega)$ \cite{aguado}:
\begin{equation}
	S_I(\omega)=\frac{1}{R_N} \left[ \frac{\hbar\omega + eV}{1-\exp{\left(-\frac{\hbar \omega+eV}{k_B T}\right)}}+ \frac{\hbar\omega - eV}{1-\exp{\left(-\frac{\hbar \omega-eV}{k_B T} \right) }}\right]
\end{equation}
This expression simplifies in the limit where $|eV|,\hbar |\omega| \gg k_B T$ and we have, with $V>0$, $S_I(\omega)= 0$ for $\hbar \omega < -eV$, $(eV+ \hbar \omega)/R_N$ for $-eV < \hbar \omega < eV$ and $2 \hbar \omega/R_N$ for $\hbar\omega > eV$.
Because $V_{Jj} > 2 \Delta$ we are in the regime $eV > \hbar \omega$. The PAT current measured is shown on Fig. \ref{fig:SNJJ}. From it one can extract the noise spectral density using equation \ref{Ipatsimple}, knowing the PAT current and the IV characteristic of the junction without noise (i.e. $I_{QP,0}(V_{SIS})$). Numerically it is easier to extract $S_V(\omega)=|Z(\omega)|^2 S_I(\omega)$ and then deduce $S_I(\omega)$ from the knowledge of the transimpedance.
The value of the noise deduced from this fit is consistent with the Poissonian value of the non-symmetrized noise $e I_{Jj}$. However the accuracy obtained numerically is not yet high enough to see clearly the frequency dependent part of the noise.

\section{Conclusion}
We have demonstrated narrow band and broad band HF detection of non-symmetrized noise of a Josephson junction using a SIS detector. This technique was used to detect the quantum noise of a charge qubit, which shows a peak at the frequency of the coherent charge oscillation (see reference \cite{deblock}). The SIS detector is operated as an on-chip spectrum analyzer and is applicable for correlation measurements on a wide range of electronic quantum devices.

\end{document}